\documentclass[a4paper, 10pt, conference]{ieeeconf}      % Use this line for a4
\IEEEoverridecommandlockouts                              % This command is only
\usepackage{amssymb}
\usepackage{graphics}
\usepackage{epsfig}
\usepackage{mathptmx}
\usepackage{times}
\usepackage{amsmath}
\usepackage{amssymb}
\usepackage{graphicx}
\usepackage{amssymb}
\usepackage{epstopdf}
\usepackage{multirow}
\usepackage{setspace}
\usepackage{amssymb}
\usepackage{amsfonts}%
\usepackage[left=54pt,top=54pt,right=37pt,bottom=104pt]{geometry}

\newtheorem{theorem}{Theorem}
\newtheorem{remark}{Remark}

\newtheorem{definition}{Definition}
\newtheorem{example}{Example}

\title{\LARGE \bf
Basic Properties and Stability of Fractional-Order Reset Control Systems
}

\author{S. Hassan HosseinNia$^{1}$ and In\'es Tejado$^{1,2}$ and Blas M. Vinagre$^{1}$ % <-this % stops a space
\thanks{*In\'{e}s Tejado would like to thank the Portuguese Funda\c{c}\~{a}o para a Ci\^{e}ncia e a Tecnologia (FCT) for the grant with reference SFRH/BPD/81106/2011. This work has been supported by the Spanish Ministry of Economy and Competitiveness under the project DPI2012-37062-C02-02 and by FCT, through IDMEC under LAETA.}% <-this % stops a space
\thanks{$^{1}$ S.H. HosseinNia, I. Tejado and B.M. Vinagre are with Department of Electrical, Electronic and Automation Engineering, Industrial Engineering School, University of Extremadura, 06006 Badajoz, Spain.
        {\tt\small e-mail: \{hoseinnia;bvinagre;itejbal\}@unex.es}}
\thanks{$^{2}$I. Tejado is also with IDMEC, Instituto Superior T\'{e}cnico, Universidade T\'ecnica de Lisboa, 1049-001 Lisbon, Portugal. {\tt\small e-mail: ines.tejado@ist.utl.pt}}
}

\begin{document}

\maketitle
\thispagestyle{empty}
\pagestyle{empty}

%%%%%%%%%%%%%%%%%%%%%%%%%%%%%%%%%%%%%%%%%%%%%%%%%%%%%%%%%%%%%%%%%%%%%%%%%%%%%%%%
\begin{abstract}
Reset control is introduced to overcome limitations of linear control. A reset controller includes a linear controller which resets some of  states to zero when their input is zero or certain non-zero values. This paper studies the application of the fractional-order Clegg integrator (FCI) and compares its performance with both the commonly used first order reset element (FORE) and traditional Clegg integrator (CI). Moreover, stability of reset control systems is generalized for the fractional-order case. Two examples are given to illustrate the application of the stability theorem.
\end{abstract}

%%%%%%%%%%%%%%%%%%%%%%%%%%%%%%%%%%%%%%%%%%%%%%%%%%%%%%%%%%%%%%%%%%%%%%%%%%%%%%%%
\section{INTRODUCTION}
Reset controllers were introduced to overcome limitations of linear controllers. For instance, in the time domain it is not possible to fulfil all characteristics and specifications --rise time, overshoot and settling time-- or in the frequency domain water-bed effect will not let the system satisfies all specifications \cite{clegg58,krishnan1974,horowitz1975,hollot2001,banos2011}. So, the main reason for using reset controllers is that, just by including the mechanism of resetting, they are able to overcome fundamental limitations in linear systems. 

The reset controller was firstly investigated by Clegg to reduce phase lag while retaining the integrator's desirable magnitude slope in the frequency response \cite{clegg58}. The Clegg integrator (CI) was introduced as a solution for improving feedback performance, due to its ability to provide the magnitude slope of a linear integrator ($-20$ dB/dec) but with a phase (about $-38^\circ$) much more favourable in terms of phase margins and robustness. More general reset structures have been proposed later to improve its performance, such as the first order reset element (FORE) controller \cite{krishnan1974,horowitz1975} and other advanced reset controllers in \cite{beker2004,zheng2007}, which allow higher order controllers and different ways of resetting.

Stability of reset control systems has received many attention in the field. Necessary and sufficient conditions for internal stability for a restricted class of systems characterized by a CI and second order plant were studied in \cite{hu1997}. Stability of reset control systems under constant inputs was analysed in \cite{beker1999,chen2000b} and its experimental application was shown in \cite{zheng2000}. BIBO stability and asymptotic tracking of FORE were established in \cite{chen2000b,chen2001}. In \cite{beker2004}, not only a testable necessary and sufficient condition for stability was given, but also links to both uniform bounded-input bounded-state stability and steady-state performance.

In what concerns the use of fractional calculus in control, the fractional-order integrator (FI) has been considered as an alternative reference system for control purposes in order to obtain closed-loop controlled systems robust to gain changes \cite{oustaloup1995,manabe1960}. From another point of view, FI can be used in feedback control in order to introduce both a constant phase lag and a magnitude slope proportional to the integration order. In other words, FI can be used with the same purposes that the reset integrator. Likewise, the fractional-order Clegg integrator (FCI) has been studied in some papers: its fundamentals can be found in \cite{Vinagre2007,Monje10}, whereas numerical values for the describing functions were reported in \cite{valerio2012}. In addition, an optimized fractional-order conditional integrator (OFOCI) was also proposed in \cite{luo2011}. 

Given this context, the purpose of this paper is to review basis of FCI and compare its performance with CI and FORE. Furthermore, stability conditions for fractional-order reset systems will be presented by generalizing some of the aforementioned methods. 

The rest of the paper is organized as follows. Section \ref{Dreset} addresses dynamics of fractional-order reset control systems. Section \ref{Rcont} recalls the main properties of CI and FORE and compares them with FCI. Stability of fractional-order reset control systems is analysed in Section \ref{QSFCI}. Finally, Section \ref{Rcon} draws the conclusions of this paper.

\section{Dynamics of Fractional-Order Reset Control Systems \label{Dreset}}

\begin{figure}[ptbh]
\begin{center}
\includegraphics[width=0.50\textwidth]{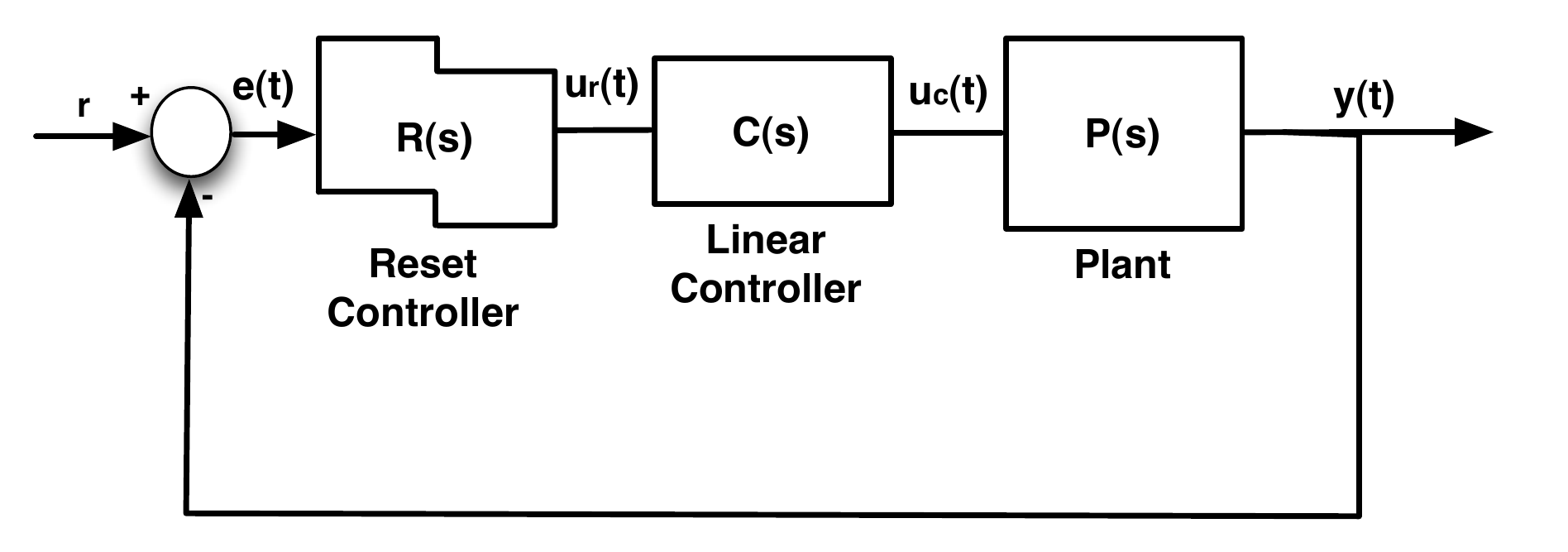}
\end{center}
\caption{Block diagram of a general reset control system}
\label{Reset_Block}%
\end{figure}

The block diagram of a general reset control system is shown in Fig. \ref{Reset_Block}. In a general form, the dynamics of the reset controller can be described by a fractional-order differential inclusion (FDI) equation as:
\begin{eqnarray}
\begin{matrix}
D^\alpha x_r(t)=A_r x_r(t)+B_re(t), \ e(t)\neq0,\\ 
x_r(t^+)=A_{R_r}x_r(t), \ e(t)=0,\\
u_r(t)=C_rx_r(t),
\end{matrix}
\label{reseteq}
\end{eqnarray}
where $0< \alpha \leq 1$ is the order of differentiation, $x_r(t) \in \mathbb{R}^{nr}$ is the reset controller state and $u_r(t) \in \mathbb{R}$ is its output. The operator $D^{\alpha}$ denotes the generalization of
the differential and integral operations whose expression, according to Gr\"{u}nwald--Letnikov definition, is given by (see e.g.~\cite{Podlubny_99a}):
\begin{eqnarray}
D^ \alpha f(t)= \lim_{h \rightarrow 0} \frac{1}{h^ \alpha}
\sum_{i=0}^{[(t-e)/h]}(-1)^i   {\alpha \choose i}f(t-ih),
\label{eq:02}
\end{eqnarray}
where $e$ and $t$ are the lower and upper bounds of the operation, $\alpha \in\mathbb{R}$ is the order and $[.]$ means the integer part.
The matrix $A_{R_r}  \in \mathbb{R}^{n_r\times n_r}$ identifies that subset of states $x_r$ that are reset (the last ${\mathcal{R}}$ states) and has the form $A_{R_r}=\begin{bmatrix}
I_{n_{\bar{\mathcal{R}}}}& 0 \\ 
0 & 0_{n_{\mathcal{R}}}
\end{bmatrix}$ with $n_{\bar{\mathcal{R}}}=n_r-n_{\mathcal{R}}$. The linear controller $C(s)$ and plant $P(s)$ have, respectively, state space representations as follows:
\begin{eqnarray}
\begin{matrix}
D^\alpha x_c(t)=A_cx_c(t)+B_cu_r(t),\\ 
u_c(t)=C_cx_c(t),
\end{matrix}
\label{Conteq}
\end{eqnarray}
and
\begin{eqnarray}
\begin{matrix}
D^\alpha x_p(t)=A_px_p(t)+B_pu_c(t),\\ 
y(t)=C_px_p(t),
\end{matrix}
\label{Syseq}
\end{eqnarray}
where $A_p\in \mathbb{R}^{n_p \times n_p}$, $B_p \in \mathbb{R}^{n_p\times 1}$, $C_p \in \mathbb{R}^{1\times n_p}$, $A_c\in \mathbb{R}^{n_c\times n_c}$, $B_c \in \mathbb{R}^{n_c\times 1}$ and $C_c \in \mathbb{R}^{1\times n_c}$. 

The closed-loop reset control system can be then described by the following FDI:
\begin{eqnarray}
\begin{matrix}
D^\alpha x(t)=A_{cl} x(t)+B_{cl}r, \ x(t) \notin  \mathcal{M}\\ 
x(t^+)=A_{R}x(t), \  x(t)\in \mathcal{M}\\
y(t)=C_{cl}x(t)
\end{matrix}
\label{CLeq}
\end{eqnarray}
where
$x=\begin{bmatrix}
x_p\\ 
x_c\\ 
x_r
\end{bmatrix}$,
$A_{cl}=\begin{bmatrix}
 A_p & B_pC_c & 0 \\ 
 0& A_c & B_cC_r \\ 
 -B_rC_p& 0 & A_r
\end{bmatrix}$,
$A_{R}=\begin{bmatrix}
 I_{n_p} & 0 & 0 \\ 
 0 & I_{n_c}& 0 \\ 
 0 & 0 & A_{R_r}
\end{bmatrix}$, $C_{cl}=\begin{bmatrix}
C_p & 0 & 0 \end{bmatrix}$ and $B_{cl}=\begin{bmatrix}
0 & 0 & B_r \end{bmatrix}^T$. The reset surface $\mathcal{M}$ is defined by:
\begin{eqnarray}
\begin{matrix}
\mathcal{M}=\left \{ x\in\mathbb{R}^n: C_{cl}x=0,\ (I-A_R)x\neq0 \right \},
\end{matrix}
\label{surf}
\end{eqnarray}
where $n=n_r+n_c+n_p$. In the case of having an integer-order controller or an integer-order system, the state space should be realized as an augmented system as follows \cite{Sierociuk2010,Hosseinnia_10a}. Consider the following integer-order system
\begin{equation}
\begin{array}{c}
\dot{x}(t)=A x(t)+Bu(t)\\ 
y(t)=Cx(t)
\end{array},
 \label{IH}
\end{equation}%
where $x\in 
%TCIMACRO{\U{211d} }%
%BeginExpansion
\mathbb{R}
%EndExpansion
^{n}$ and $y(t)$ are the state vector and the output of the system. The integer-order state space model can be rewritten in the following augmented
fractional-order system:%
\begin{equation}
\begin{array}{c}
D^{\alpha}\mathcal{X}(t)=\mathbf{A}\mathcal{X}(t)+\mathbf{B}u(t) \\ 
y(t)=\mathbf{C}\mathcal{X}(t)
\end{array}, 
\label{AFH} 
\end{equation}
\begin{equation*}
\mathbf{A} =\left[ 
\begin{array}{cccccc}
0 & I & 0 & \cdots  & 0 & 0 \\ 
0 & 0 & I & \cdots  & 0 & 0 \\ 
\vdots  & \vdots  & \vdots  & \vdots  & \vdots  & \vdots  \\ 
0 & 0 & 0 & \cdots  & 0 & I \\ 
A & 0 & 0 & \cdots  & 0 & 0%
\end{array}%
\right] ,\ \mathbf{B=}\left[ 
\begin{array}{c}
0 \\ 
0 \\ 
\vdots  \\ 
0 \\ 
B%
\end{array}%
\right] ,
\end{equation*}
\begin{equation}
\\ \mathbf{C} =\left[ 
\begin{array}{cccccc}
C & 0 & 0 & \cdots  & 0 & 0%
\end{array}
\right] ,  
\end{equation}
where $\mathcal{X}=\left[ x\
x_{a,1}\ ...\ x_{a,p-1}\right] ^{T}$ is the vector of augmented states, $p=\frac{1}{\alpha}$, and $I$ is the identity matrix.

\section{Properties of Reset Controllers \label{Rcont}}

\begin{figure}[ptbh]
\begin{center}
\includegraphics[width=0.50\textwidth]{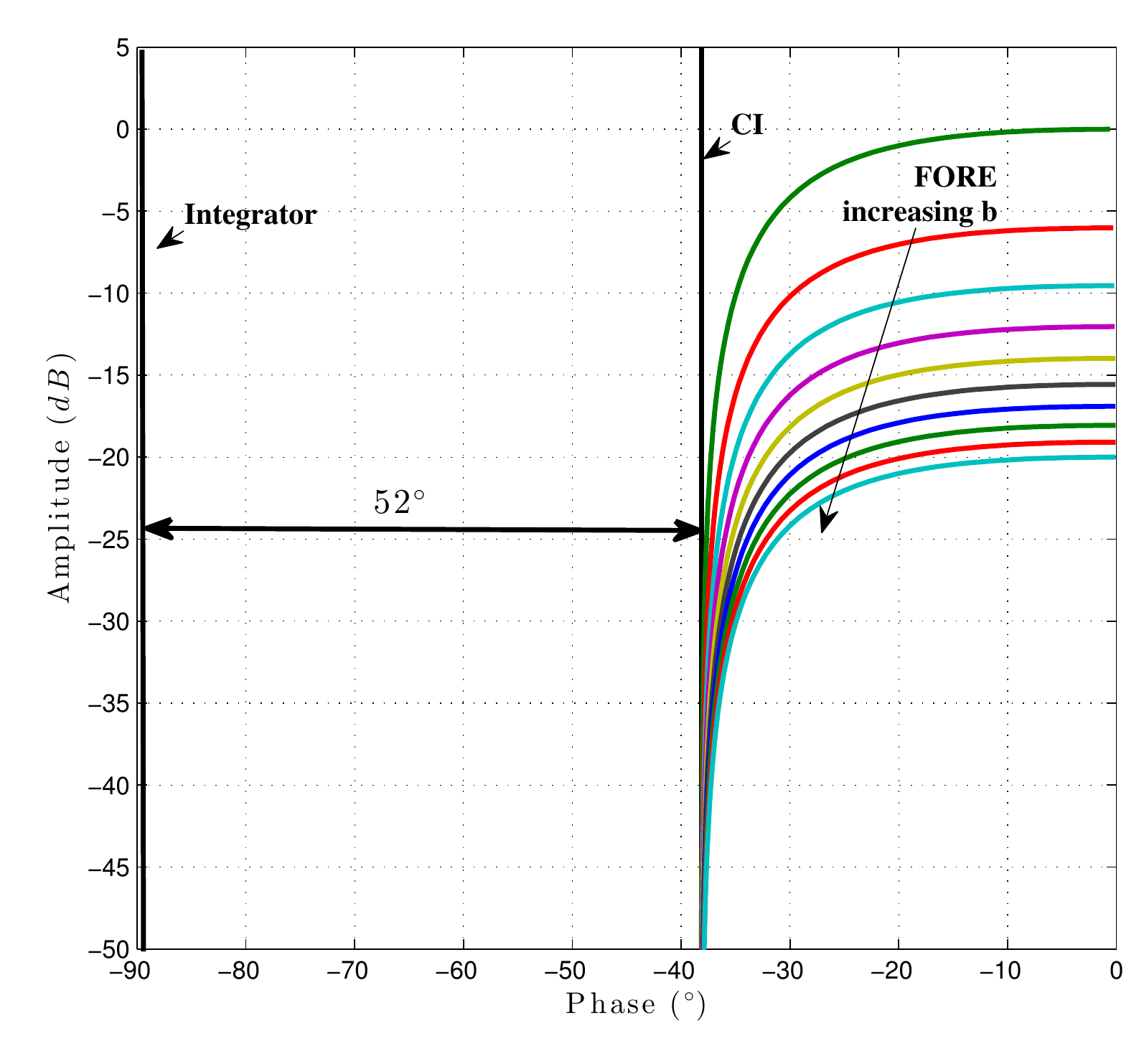}
\end{center}
\caption{Nichols chart of FORE and CI with respect to the classical integrator}
\label{FORE_DFN}%
\end{figure}

For a given system its describing function (DF) can be defined calculated by:
\begin{equation}
N(A,\omega)=\frac{2j\omega}{\pi A}\int_0^\pi{y(t)e^{-j\omega t}dt}.
\label{DF}
\end{equation}
FORE is a simple reset compensator with a first order base compensator given by
\begin{equation} 
FORE(s)=\frac{K}{s+b}.
\end{equation} 
Applying (\ref{DF}), its DF can be given by \cite{horowitz1975,banos2011}:
\begin{equation}
N(A,\omega)_{FORE}=\frac{K}{b+j\omega}\left(1+j \frac{2\omega^2 \left( 1+e^{-b\frac{\pi}{\omega}}\right)}{\pi\left( b^2+\omega^2 \right)} \right).
\label{DF_FORE}
\end{equation}
Thus, the DF of CI can be obtained by substituting $b=0$ and $K=1$ in (\ref{DF_FORE}), which yields:
\begin{equation}
N(A,\omega)_{CI}=\frac{4}{\pi\omega}\left(1-j \frac{\pi}{4} \right),
\label{DF_CI}
\end{equation}
Therefore, it is clear that CI gives a phase lead of almost $52^\circ$ with respect to a classical integrator (it also increases the gain by a factor of about $1.62$). Figure \ref{FORE_DFN} shows this fundamental property of CI and FORE by means of the Nichols chart.

On the contrary, it has been shown that FCI has a tunable phase lag and its DF can be represented as \cite{Vinagre2007}:
\begin{equation}
N(A,\omega)_{FCI}=\frac{4}{\pi\omega^\alpha}\left(\sin \left(\alpha\frac{ \pi}{2}\right)+\frac{\pi}{4}e^{-j\alpha \frac{\pi}{2}} \right),
\label{DF_FCI}
\end{equation}

\begin{figure}[ptbh]
\begin{center}
\includegraphics[width=0.50\textwidth]{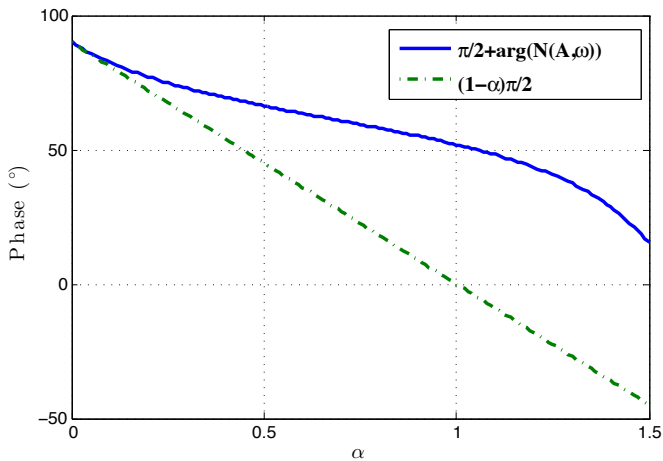}
\end{center}
\caption{Phase difference between both the FCI and the FI in comparison with integer-order linear integrator (II). (FCI \textit{vs} II: $\frac{\pi}{2}+\arg(N(A,\omega)_{FCI}))$, FI \textit{vs} II: $(1-\alpha)\frac{\pi}{2}$)}
\label{DF_FCI2}
\end{figure}

Figure~\ref{DF_FCI2} compares the phase difference between both the FCI and the FI in comparison with the integer-order linear integrator (II) for different values of the order $\alpha$. As observe, the phase lag depends on the value of $\alpha$ for both cases, but is always higher when using the FCI for $\alpha<1$. In particular, when $\alpha=1$, the phase difference between the FCI and the II is about $52^\circ$ (actually, the FCI is the CI) and $0$ for the other case. Note that this phase difference can be considered as the phase margin to be added to the system. As an example, CI cannot compensate 60$^\circ$ but this specification can be achieved by an FCI of order $\alpha=0.5$, namely FCI$^{0.5}$, or by an FI of order $\alpha=0.26$, i.e., FI$^{0.26}$. As observe, FCI has higher order than the FI which leads the system to have a faster response. Similarly, if one needs to compensate about 52$^\circ$, it can be possible using CI which causes faster response than an FI$^{0.42}$. Although the FCI and, in some cases, the CI have better performance than the FI, their applicability depends on the system dynamics and the particular application. In the following example all these reset strategies will be compared.

\begin{example} Comparison of different reset controllers to reduce the overshoot
\label{Rest_exp}
\end{example}
One of the motivation of using reset control is to reduce the overshoot in a step response. For example, let us consider the same feedback system as in \cite{hollot2001} whose transfer function and controller are
\begin{equation}
P(s)=\frac{1}{s^2+0.2s},
\label{SOP}
\end{equation}
and
\begin{equation}
C(s)=s+1,
\label{Cont_SOP}
\end{equation}
respectively. The system shows an overshoot of $70\%$ and, consequently, the aim is to design the different reset controllers to reduce it, obtaining faster response at the same time. In \cite{hollot2001}, the authors used a FORE with $b=1$ and reduced the overshoot to about $40\%$. Next, this controller is going to be compared with CI, FI and FCI --both fractional controllers of order $\alpha=0.5$--.

\begin{figure}[ptbh]
\begin{center}
\includegraphics[width=0.450\textwidth]{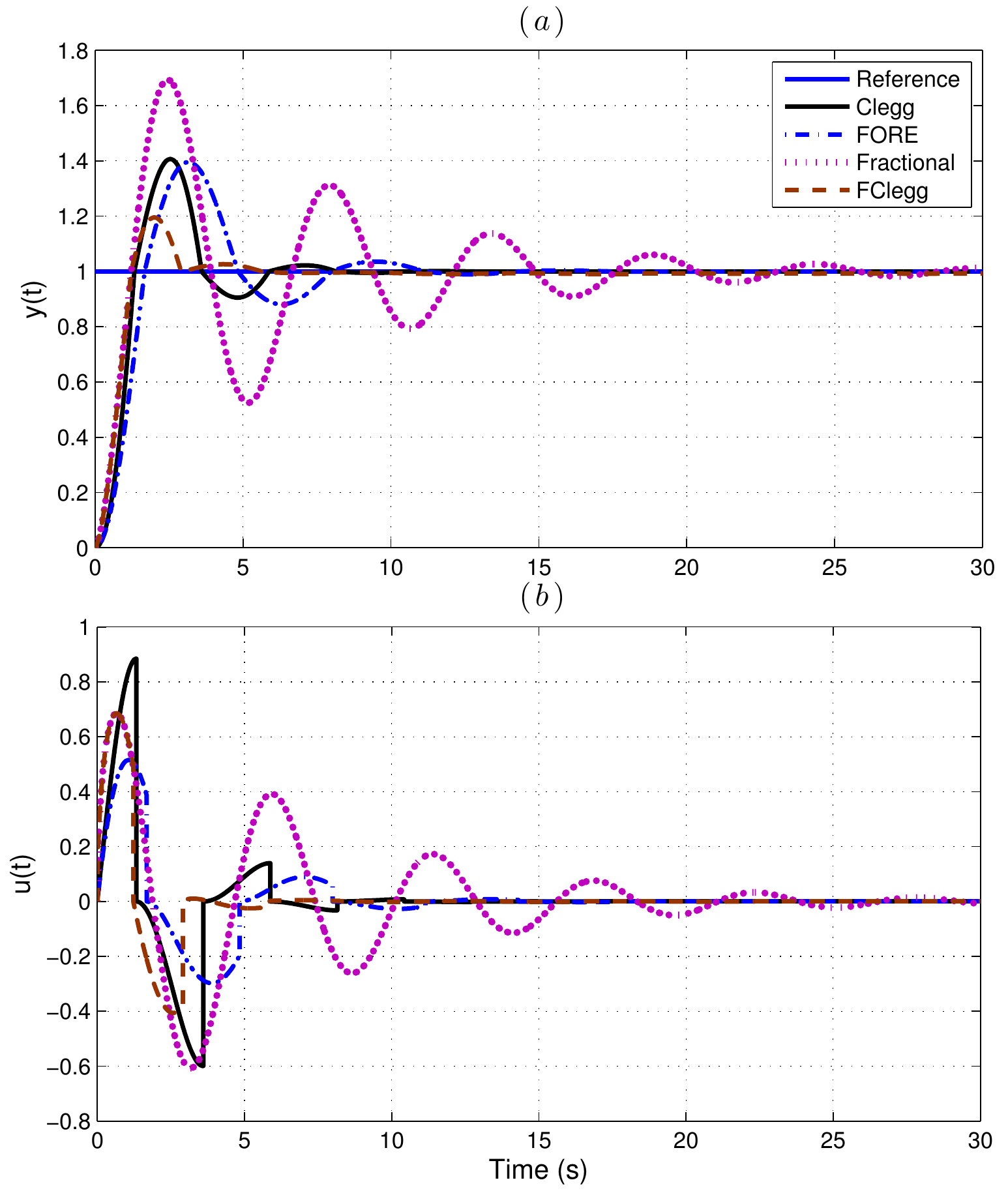}
\end{center}
\caption{Responses of system (\ref{SOP}) using different reset controllers: ($a$) Outputs ($b$) Control efforts}
\label{Compare_reset}
\end{figure}

The simulation results are plotted in Fig. \ref{Compare_reset}. It can be seen that the system response applying CI has an overshoot of $41\%$, a bit higher than the obtained by FORE but the response is faster. As expected, the system performance when using FI is poor --it corresponds to the worst response--. On the contrary, the best response is obtained by FCI, which is capable of reducing the overshoot to about $19\%$. It should be also commented that there exist other ways to reduce the overshoot but it may cause the limitation in the response and the aim in this particular example was obtaining faster response lower overshoot at the same time.

\section{Stability of Fractional-Order Reset Control Systems \label{QSFCI}}

This section concerns stability of fractional-order reset control systems. Firstly, some definitions needed to our main results are given. 

\subsection{Preliminaries}

Consider a fractional-order linear time invariant (FO-LTI) system as:
\begin{equation}
D^{\alpha}x(t)=Ax(t), x\in\mathbb{R}^{n}
\label{FOLTI}.
\end{equation}
In particular, $t^{-a}$ stability will be used to refer to asymptotic stability of
fractional-order systems. 

\begin{definition}
[$t^{-a }$ stability \cite{Moze_07}] The trajectory $x(t)=0$ of system (\ref{FOLTI}) is $t^{-a}$ asymptotically stable if the
uniform asymptotic stability condition is met and if there is a positive real
$a$ such that:\newline$\forall\left\Vert x\left(  t\right)  \right\Vert
,t\leq t_{0}\ \exists\ N\left(  x\left(  t\right)  ,t\leq t_{0}\right)
,\ t_{1}\left(  x\left(  t\right)  ,t\leq t_{0}\right)$ such that $\forall
t\leq t_{0},\ \left\Vert x\left(  t\right)  \right\Vert \leq\ N\left(
t-t_{1}\right)  ^{-a }.$
\end{definition}

\begin{theorem}[\cite{Moze_07} ]
A fractional-order system given by (\ref{FOLTI}) with order $\alpha$,
$0<\alpha\leq1$, is $t^{-a }$ asymptotically stable if and
only if there exists a positive definite matrix $P\in\mathbb{R}^{n}$ such that
\begin{equation}
\left(  -\left(  -A\right)  ^{\frac{1}{2-\alpha}}\right)  ^{T}P+P\left(
-\left( -A\right)  ^{\frac{1}{2-\alpha}}\right)  \leq0.
\end{equation}
\label{Moze0}
\end{theorem}

\begin{theorem}[Lyapunov-like theorem \protect\cite{Goebel_09}\label{HLS}]
Consider a closed-loop reset system given by (\ref{CLeq}). If there exists a
Lyapunov-function candidate $V(x)$ such that
\begin{eqnarray}
\dot{V}(x)<0, \ x(t) \notin  \mathcal{M},
\label{LLC1}
\end{eqnarray}
\begin{eqnarray}
\bigtriangleup V(x)=V(x(t^+))-V(x(t))\leq0, \ x(t) \in  \mathcal{M},
\label{LLC2}
\end{eqnarray}
then there exists a left-continuous function $x(t)$ satisfying (\ref{CLeq}) for all $t\geq0$,
and the equilibrium point $x_e$ is globally uniformly asymptotically stable.
\label{TLL}
\end{theorem}

\begin{definition} Reset control system (\ref{CLeq}) is said to satisfy the H$_\beta$-condition if there exists a $\beta \in \mathbb{R}^{n_{\mathcal{R}}}$ and a positive-definite matrix $P_{\mathcal{R}} \in \mathbb{R}^{n_{\mathcal{R}} \times n_{\mathcal{R}}}$ such that
\begin{eqnarray}
H_\beta(s)=\begin{bmatrix}
\beta C_p & 0_{n_{\bar{\mathcal{R}}}} & P_{\mathcal{R}}
\end{bmatrix}\left ( sI-\mathcal{A} \right )^{-1}\begin{bmatrix}
0\\ 
0^T_{\bar{\mathcal{R}}}\\ 
I_{\mathcal{R}}
\end{bmatrix},
\label{Hb}
\end{eqnarray}
where $\mathcal{A}= \left( -\left(-A_{cl}\right)  ^{\frac{1}{2-\alpha}}\right)$.
\end{definition} 

\subsection{Asymptotic Stability}

According to \cite{beker1999,beker2004,chen2000}, an integer-order reset control system of the form of (\ref{CLeq}) --with $\alpha=1$-- is asymptotically stable if and only if it satisfies the H$_\beta$-condition. The same idea can be used to prove the stability of fractional-order reset systems.

Now, consider $V(z(t))=z(t)^T\mathcal{P}z(t),\ \mathcal{P}\in\mathbb{R}^{N\times N}$ as a Lyapunov candidate for the unforced reset system (\ref{CLeq}) ($r=0$) where $x=[0,\cdots,0,1]z(t),\ z(t)\in\mathbb{R}^{N\times N}, \ \dot{z}=A_{f}z(t),$ and $A_{f}=
\begin{bmatrix}
0 & \cdots & 0 & A^{1/\alpha}\\
A^{1/\alpha} & \cdots & 0 & 0\\
& \ddots &  & \vdots\\
0 &  & A^{1/\alpha} & 0
\end{bmatrix}$ (see \cite{Moze_07} more details for this transformation). Then, in accordance with \cite{Moze_07}, the necessary and sufficient condition to satisfy $\dot{V}(z(t))<0$ when $\frac{2}{3}<\alpha\leq1$ is:
\begin{eqnarray}
\nonumber
\left(A^\frac{1}{\alpha}\right) ^{T}P+P\left(A^\frac{1}{\alpha}\right) <0, \ x(t) \notin \mathcal{M}.
\end{eqnarray}
where $P(\subset\mathcal{P})\in\mathbb{R}^{n\times n}>0$. Likewise, based on results stated in Theorem \ref{Moze0}, the necessary and sufficient condition for $0<\alpha\leq1$ is
\begin{eqnarray}
\nonumber
\mathcal{A} ^{T}P+P\mathcal{A}  <0, \ x(t) \notin \mathcal{M}.
\label{unf}
\end{eqnarray}
Transforming the second equation of reset system (\ref{CLeq}), we have
\begin{eqnarray}
z(t^+)=\begin{bmatrix}
I_{N-n}& 0\\
0 & A_R
\end{bmatrix}z(t),
\label{disnew}
\end{eqnarray}
where $I_{N-n}$ is identity matrix with dimension of ${N-n}$. Thus, $\bigtriangleup V(z(t))<0$ if 
\begin{equation*}
V(z(t^+))-V(z(t))=
\end{equation*}
\begin{equation}
z^T(t)\left(\begin{bmatrix}
I_{N-n}& 0\\
0 & A^T_R
\end{bmatrix}\mathcal{P}+\mathcal{P}\begin{bmatrix}
I_{N-n}& 0\\
0 & A_R
\end{bmatrix}\right)z(t)\leq0.
\label{deltaz}
\end{equation}
Then, (\ref{deltaz}) is satisfied if $V(x(t^+))-V(x(t))\leq0$,
\begin{equation*}
x^T(t)(A_R^TPA_R-P\leq0)x(t)\leq0, \ x(t) \in  \mathcal{M}.
\end{equation*}
Therefore, Theorem \ref{TLL} can be reshaped in the following remark.

\begin{remark}
Choosing $V(z)=z(t)^T\mathcal{P}z(t),\ \mathcal{P}\in\mathbb{R}^{N\times N}$ as a Lyapunov candidate, and applying Theorem \ref{Moze0}, fractional-order reset system (\ref{CLeq}) is asymptotically stable if and only if:
\begin{eqnarray}
\label{Fcr1}
\mathcal{A} ^{T}P+P\mathcal{A}  <0, \ x(t) \notin  \mathcal{M},\\
\label{Fcr2}
A_R^TPA_R-P\leq0, \ x(t) \in  \mathcal{M}.
\end{eqnarray}
\label{remarkres1}
\end{remark}
Consider a reset system with constant input and let us define $\bold{x}(t)=x(t)-x_e=x(t)+A_{cl}^{-1}B_{cl}r$. Thus, reset system (\ref{CLeq}) can be rewritten as:
\begin{eqnarray}
\begin{matrix}
D^\alpha \bold{x}(t)=A_{cl} \bold{x}(t), \ x(t) \notin  \mathcal{M},\ x(0)=x_0\\ 
\bold{x}(t^+)=A_{R}(\bold{x}(t)+x_e), \  x(t)\in \mathcal{M}\\
y(t)=C_{cl}x(t).
\end{matrix}
\label{CLeq2}
\end{eqnarray}
Choosing a similar Lyapunov function, i.e, $V(z(t))=z(t)^T\mathcal{P}z(t),\ \bold{x}(t)=[0,\cdots,0,1]z(t)$, system (\ref{CLeq2}) is stable if conditions (\ref{LLC1}) and (\ref{LLC2}) are satisfied. Comparing (\ref{CLeq}) and (\ref{CLeq2}), condition (\ref{LLC1}) is fulfilled if (\ref{Fcr1}) is satisfied, and similarly to the unforced system $\bigtriangleup V(z(t))\leq0$ if $\bigtriangleup V(\bold{x}(t))\leq0$ (see (\ref{disnew}) and (\ref{deltaz})). Thus,
\begin{equation*}
\bigtriangleup V(\bold{x}(t))=V(\bold{x}(t^+))-V(\bold{x}(t))=
\end{equation*}
\begin{equation*}
(\bold{x}(t)+x_e)^TA_R^TPA_R(\bold{x}(t)+x_e)- \bold{x}(t)^TP\bold{x}(t)<0 \ \rightarrow 
\end{equation*}
\begin{equation*}
\bold{x}^T(t)(A_R^TPA_R-P)\bold{x}(t)<-(M=x_e^TA_R^TPA_Rx_e) \ \rightarrow 
\end{equation*}
\begin{equation*}
\bold{x}(t)^T\left(\left(A_R^TPA_R-P\right)<0\right)\bold{x}(t)<0.
\end{equation*}
Therefore, Remark \ref{remarkres1} will be also applicable to this special case.
Define $\tilde{\mathcal{M}}=\left \{ x\in\mathbb{R}^n: C_{cl}x(t)=r \right \}$, and let $\Phi$ be a matrix whose columns span $\tilde{\mathcal{M}}$. Since $\tilde{\mathcal{M}} \subset  \mathcal{M}$, (\ref{Fcr2}) is implied by
\begin{eqnarray}
\label{Fcr}
\Phi \left (A_R^TPA_R-P<0 \right ) \Phi\leq0.
\end{eqnarray}

A straightforward computation shows that inequality (\ref{Fcr}) holds for some positive-definite symmetric matrix $P$ if there exists a $\beta \in \mathbb{R}^{n_{\mathcal{R}}}$ and a positive-definite $P_{\mathcal{R}} \in \mathbb{R}^{n_{\mathcal{R}} \times n_{\mathcal{R}}} $ such that
\begin{eqnarray}
\begin{bmatrix}
0 & 0_{\bar{\mathcal{R}}} & I_{\mathcal{R}}
\end{bmatrix} P
=\begin{bmatrix} \beta C_p & 0_{n_{\bar{\mathcal{R}}}} & P_{\mathcal{R}}
\end{bmatrix}.
\label{Hbeta}
\end{eqnarray}
To analyze stability, it suffices to find a positive-definite symmetric matrix $P$ such that (\ref{Fcr1}) and (\ref{Hbeta}) hold. Taking into account Kalman-Yakubovich-Popov (KYP) lemma \cite{slotine1991}, such $P$ exists if H$_\beta(s)$ in (\ref{Hb}) is strictly positive real (SPR) for some $\beta$. In addition, in accordance with \cite{ioannou1987}, it is obvious that the H$_\beta(s)$ is SPR if 
\begin{eqnarray}
\label{FrSPR}
\left |\arg(H_\beta(j\omega)) \right |<\frac{\pi}{2}, \forall \omega.
\end{eqnarray}
Therefore, these results can be stated in the following theorem.

\begin{theorem} The closed-loop fractional-order reset control system (\ref{CLeq}) is asymptotically stable if and only if it satisfies the H$_\beta$-condition (\ref{Hb}) or its phase equivalence (\ref{FrSPR}).
\label{FRQS}
\end{theorem}

In order to show the applicability of the aforementioned results, two examples are given next.

\begin{example} Stability analysis of a fractional-order system controlled by FCI
\label{FCIexpI}
\end{example}
Let us consider a plant, $P(s)=1/s$, controlled by an FCI$^{0.5}$ in negative feedback without exogenous inputs. Therefore, the closed-loop system can be represented by in augmented state space form as follows (see \cite{Hosseinnia_10a})
\begin{eqnarray*}
D^{0.5}x_p(t)=\begin{bmatrix}
0 & 1 \\
0 & 0
\end{bmatrix}x_p(t).
\end{eqnarray*}
If the state vector is $x(t)=(x_p(t), x_r(t))^T$ with $x_p(t)=(x_{p_1}(t),x_{p_2}(t))^T$ being the plant state and $x_r(t)$, the (reset) controller state, then it results in a reset system like that in (\ref{CLeq}) with
\begin{eqnarray*}
A_{cl}=\begin{bmatrix}
0 & 1 & 0\\
0 & 0 & 1\\
 -1 & 0 & 0 
\end{bmatrix}, \ A_R=\begin{bmatrix}
1 & 0 & 0\\
0 & 1 & 0\\
 0 & 0 & 0 
\end{bmatrix}, \ C_{cl}=\begin{bmatrix}
1 & 0 & 0
\end{bmatrix}.
\end{eqnarray*}
In addition, from (\ref{Hb}), H$_\beta$ is simply given by (for this case $n_{\mathcal{R}}=1$ and then $P_{\mathcal{R}}=1$ without loss of generality):
\begin{equation*}
H_\beta(s)=\begin{bmatrix}
\beta & 0 & 1
\end{bmatrix} \begin{bmatrix}
s+0.45 & -0.84 & -0.29 \\ 
 0.29 & s+0.45 & -0.84 \\ 
 0.84 & 0.29 & s+0.45 
\end{bmatrix}^{-1} \begin{bmatrix}
0\\ 
0\\ 
1
\end{bmatrix}
\end{equation*}
\begin{equation*}
=\frac{(s^2+0.9s+0.45)+\beta(0.29s+0.84)}{s^3+1.35s^2+1.35s+1}.
\label{pol}
\end{equation*}

\begin{figure}[ptbh]
\begin{center}
\includegraphics[width=0.50\textwidth]{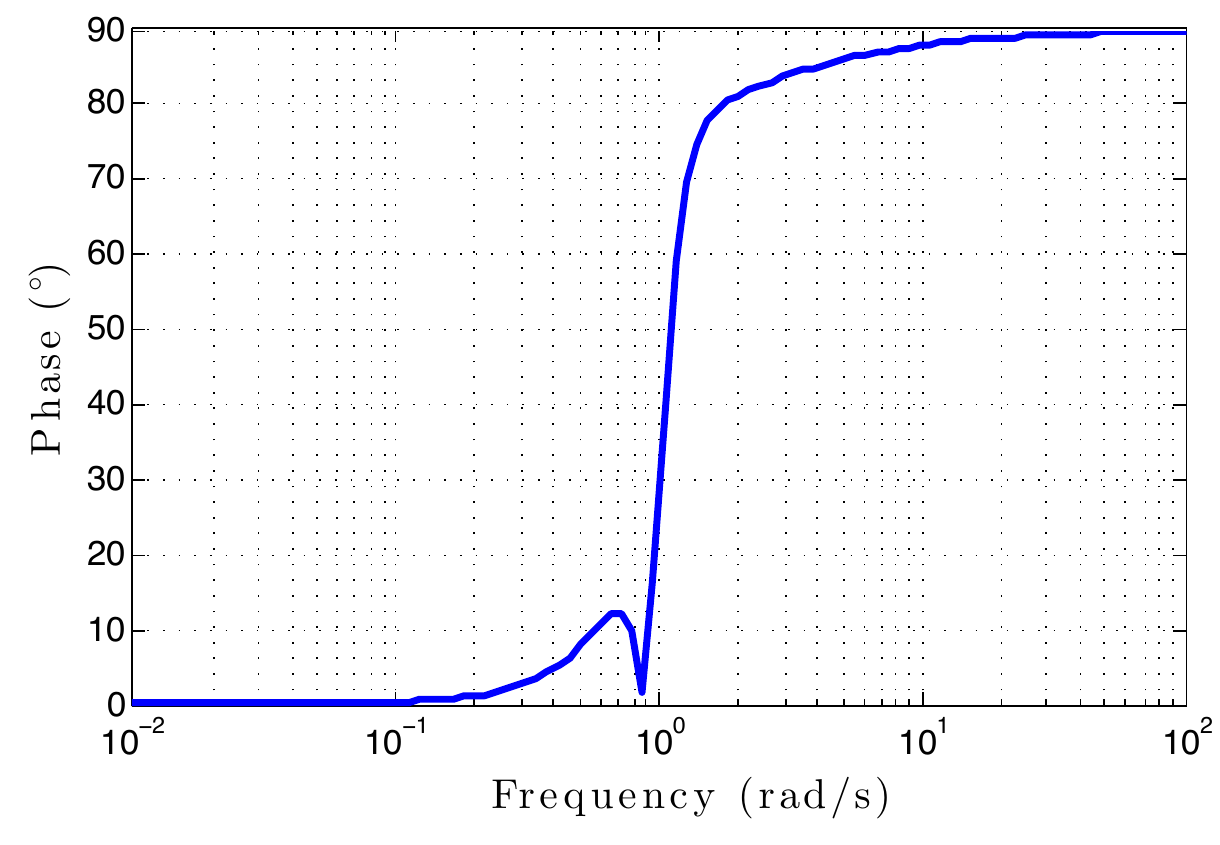}
\end{center}
\caption{Phase equivalence of H$_\beta$ (\ref{pol}) in Example \ref{FCIexpI}}
\label{figFHb}%
\end{figure}

Finally, $Re\left ( H_\beta(j\omega) \right )>0, \forall \omega>0$ for $-0.53 \leq \beta \leq 0.79$, which means that the system is SPR. The phase equivalence of (\ref{pol}) is shown in Fig. \ref{figFHb}. As observe, $\left |\arg(H_\beta(j\omega)) \right |<\frac{\pi}{2}$ for all finite $\omega >0$ and $\beta=0.3$, which proves the stability of fractional-order reset system studied in this example.

\begin{figure}[ptbh]
\begin{center}
\includegraphics[width=0.450\textwidth]{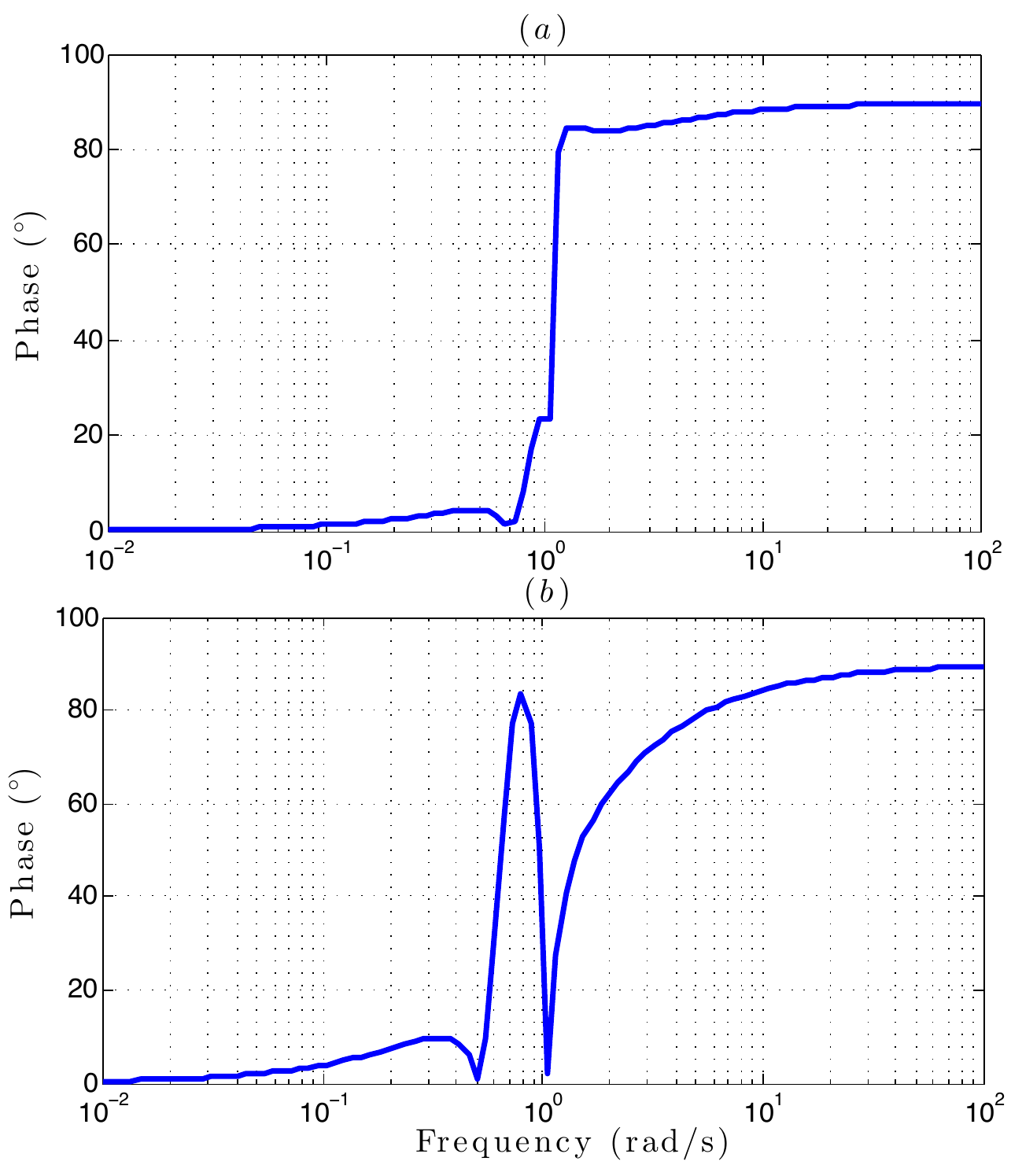}
\end{center}
\caption{Phase equivalence of H$_\beta$ in Example \ref{Rest_exp_stab}: ($a$) Applying FCI ($b$) Applying FORE}
\label{HB_FCIexp2}%
\end{figure}

\begin{example} Stability analysis of Example \ref{Rest_exp}
\label{Rest_exp_stab}
\end{example}
Let us go back to Example \ref{Rest_exp} and analyze the stability of the system when applying FORE, CI and FCI. For FORE controller, the integer-order closed-loop system can be given by:
\begin{eqnarray*}
\left\{
\begin{matrix}
\dot{x}=A_{cl}x=\begin{bmatrix}
0 & 1 & 0\\
0 & -0.2 & 1\\
 -1 & -1 & -b 
\end{bmatrix}x(t)\\ 
x(t^+)=A_Rx=\begin{bmatrix}
1 & 0 & 0\\
0 & 1 & 0\\
 0 & 0 & 0 
\end{bmatrix}x(t)\\ 
y=C_{cl}x=\begin{bmatrix}
1 & 1 & 0
\end{bmatrix}x(t)
\end{matrix}
\right.
\end{eqnarray*}
where $x(t)=\left[x_{p_1}(t),x_{p_2}(t),x_r(t)\right]^T$. And, the closed-loop system using FCI can be stated as
\begin{eqnarray*}
\left\{
\begin{matrix}
D^{0.5}\mathcal{X}(t)=\mathbf{A} _{cl}\mathcal{X}(t)=\begin{bmatrix}
0 & 1 & 0 & 0 & 0\\
0 & 0 & 1 & 0 & 0\\
0 & 0 & 0 & 1 & 0\\
0 & 0 & -0.2 & 0 & 1\\
 -1 & 0 & -1 & 0 & 0
\end{bmatrix}\mathcal{X}(t)\\ 
\mathcal{X}(t^+)=\mathbf{A}_R\mathcal{X}(t)=\begin{bmatrix}
I_4 & 0_{4,1} \\
 0_{1,4} & 0 
\end{bmatrix}\mathcal{X}(t)\\ 
y=\mathbf{C}_{cl}\mathcal{X}(t)=\begin{bmatrix}
 1 & 0 & 1 & 0 & 0
\end{bmatrix}\mathcal{X}(t)
\end{matrix}
\right.
\end{eqnarray*}
where $\mathcal{X}(t)=\left[\mathcal{X}_{p_1}(t),\cdots,\mathcal{X}_{p_4}(t),x_r(t)\right]^T$, $\mathcal{X}_{p_1}(t)=x_{p_1}(t)$, $\mathcal{X}_{p_3}(t)=x_{p_2}(t)$. According to condition (\ref{Hb}), H$_\beta$ corresponding to FORE and FCI are  simply given by (for both case FORE and FCI $n_{\mathcal{R}}=1$ and then $P_{\mathcal{R}}=1$):
\begin{equation*}
\nonumber
H^{FORE}_\beta(s)= \begin{bmatrix}
\beta & 0 & 1
\end{bmatrix}\left ( sI-A_{cl} \right )^{-1}\begin{bmatrix}
0\\ 
0\\ 
1
\end{bmatrix}=
\end{equation*}
\begin{equation}
\frac{s^2+0.2s+0.8\beta}{s^3+(b+0.2)s^2+(1+0.2b)s+1},
\label{HB_FORE_CI}
\end{equation}
and
\begin{equation}
H^{FCI}_\beta(s)=
\begin{bmatrix}
\beta & 0 & \beta & 0 & 1
\end{bmatrix}\left ( sI-\left( -\left(-\mathbf{A}_{cl}\right)  ^{\frac{2}{3}}\right) \right )^{-1}\begin{bmatrix}
0\\ 
0\\ 
0\\ 
0\\ 
1
\end{bmatrix}.
\label{HB_FCI}
\end{equation}
Therefore, using Theorem \ref{FRQS}, the closed-loop systems controlled by FORE and FCI are asymptotically stable if H$^{FORE}_\beta(s)$ and H$^{FCI}_\beta(s)$ are SPR. Substituting $b=1$ in (\ref{HB_FORE_CI}), the FORE reset system is asymptotically stable for all $0.42<\beta \leq1.46$. With respect to CI (similarly to FORE with $b=0$), stability cannot be guaranteed with this theorem. And applying FCI, it can be easily stated that the system is asymptotically stable for $\beta \leq 0.62$. In addition, the phase equivalences corresponding to (\ref{HB_FORE_CI}) and (\ref{HB_FCI}) are shown in Fig. \ref{HB_FCIexp2} for $\beta=0.5$ and $b=1$. It can be seen that both phases verifies condition (\ref{FrSPR}), which has concordance with the theoretical results.

\section{Conclusions \label{Rcon}}
In this paper, some traditional reset control strategies were compared with the fractional-order Clegg integrator (FCI). It has been demonstrated that the FCI has better performance in compensating the lead phase. Likewise, it has been shown that FCI may be capable of reducing the overshoot in a proper and better way. Lyapunov stability has been generalized for fractional-order reset systems, presenting its phase equivalence in the frequency domain. The results have shown the applicability of the proposed method to prove the stability of such fractional-order systems.

The application and stability analysis of fractional-order FORE with the base transfer function F$_r$FORE$(s)=\frac{K}{s^\alpha+b}$ will be studied as future works.

\bibliographystyle{IEEEtran}
\bibliography{Bibliography}

\end{document}